\documentclass[doublespacing,final]{elsart}
\usepackage{amssymb}
\usepackage{amsfonts}
\usepackage{graphicx}

\input{tcilatex}

\begin{document}

\begin{frontmatter}



\title{Exact nonparametric inference for detection of nonlinearity}


\author{Xiaodong~Luo\corauthref{luo}}$^\dag$ $^\ddag$, \corauth[luo]{Corresponding author}\ead{luox@maths.ox.ac.uk}
\author{Jie~Zhang}$^\dag$, \author{Michael~Small}$^\dag$, \and \author{Irene~Moroz}$^\ddag$

\begin{center}

$^\dag$ \textit{\small Department of Electronic and Information
Engineering, The Hong Kong Polytechnic University, Hung Hom, Hong
Kong}

$^\ddag$ \textit{\small Oxford Centre for Industrial and Applied Mathematics, Mathematical Institute, University of
Oxford, Oxford, UK}

\end{center}

\begin{abstract}
We propose an exact nonparametric inference scheme for the
detection of nonlinearity. The essential fact utilized in our
scheme is that, for a linear stochastic process with jointly
symmetric innovations, its ordinary least square (OLS) linear
prediction error is symmetric about zero. Based on this viewpoint,
a class of linear signed rank statistics, e.g. the Wilcoxon signed
rank statistic, can be derived with the known null distributions
from the prediction error. Thus one of the advantages of our
scheme is that, it can provide exact confidence levels for our
null hypothesis tests. Furthermore, the exactness is applicable
for finite samples. We demonstrate the test power of this
statistic through several examples.

\end{abstract}

\end{frontmatter}

\section{Introduction}

Nonlinear statistics such as correlation dimension and Lyapunov exponent 
\cite{kantz nonlinear} have been widely adopted in many fields to identify
the underlying dynamical systems. However, even for a linear stochastic
process with simple autocorrelation, these statistics could have finite and
predictable values \cite{osborne finite}, thus without careful treatment,
one may mistake a linear stochastic process for a nonlinear deterministic
system when simply examining whether these statistics are convergent. This
observation requires us to examine the basic properties of the underlying
system in order to apply nonlinear analysis methods with greater confidence.
Based on this viewpoint, to discriminate (stationary) linear stochastic
processes from nonlinear deterministic systems, various methods have been
developed, for example, the idea to investigate the orientations of the
tangents to the system trajectory within given regions \cite{kaplan direct},
the proposal to test the continuity of the underlying systems \cite{kaplan
exceptional}, and the suggestion to measure the mutual information and
redundancy within the framework of information theory \cite{palus
information}, to name but a few (see \cite{schreiber discrimination} for an
extensive study).

In general, these methods focus on exploring the difference of some
characteristic behaviors or properties between linear stochastic and
nonlinear (deterministic) systems. In addition, a confidence level is
usually preferred in order to indicate the reliability of the results. In
practical situations, if only a scalar time series from an unknown source is
available, the conventional approach for inference of the underlying system
with the confidence level, as proposed by Theiler \textit{et al. }\cite%
{theiler testing}\textit{,} is to first assign a null hypothesis to the
underlying system, and then apply the bootstrap method to produce a set of
constrained-realization surrogate data, which should have the same statistic
distribution as the original time series under the null hypothesis. Hence
based on the empirical distribution of the adopted statistic of the
surrogates and the original time series, one then determines whether to
reject the null or not. However, since the exact knowledge of the statistic
distribution is often not available, one will resort to certain
discriminating criterion to help make the decision more objectively and
determine the corresponding confidence level (if to reject). The popular
discriminating criteria appearing in the literature of nonlinear (chaotic)
time series analysis usually include two classes: parametric and
nonparametric.

The parametric criterion assumes that the statistic follows a Gaussian
distribution, and the distribution parameters, i.e. the mean and the
variance, are estimated from the finite samples. One can determine whether
to reject the null by examining whether the statistic of the original time
series follows the statistic distribution of the surrogates. The
corresponding confidence level of inference can be calculated from the
estimated statistic distribution (see our discussions later); The
nonparametric criterion \cite{theiler using} examines the ranks of the
statistic values of the original time series and its surrogates. Supposes
that the statistic of the original time series is $s_{0}$ and the surrogate
values are $\{s_{i}\}_{i=1}^{N}$ given $N$ surrogate realizations. Then if
the statistic of both the original time series and the surrogates follows
the same distribution, the probability is $1/(N+1)$ for $s_{0}$ to be the
smallest or largest among all of the values $\{s_{0},s_{1}...s_{N}\}$. Thus
if $N$ is large, when one finds that $s_{0}$ is smaller or larger than all
of the values in $\{s_{i}\}_{i=1}^{N}$, it is likely that $s_{0}$ instead
follows a different distribution from that of $\{s_{i}\}_{i=1}^{N}$. Hence
the criterion rejects the null hypothesis whenever the original statistic $%
s_{0}$ is the smallest or largest among $\{s_{0},s_{1}...s_{N}\}$, the false
rejection rate is considered as $1/(N+1)$ for one-sided tests and $2/(N+1)$
for two-sided ones.

Note that the above two simple criteria are often adopted thanks to the
difficulty in calculating the exact distribution of the nonlinear statistic
under test. Although these two criteria are heuristic, they are often
questionable in practice. For example, for the first criterion, the
normality assumption may not approximate the actual distribution well. For
the second, suppose that $s_{0}$ and $\{s_{i}\}_{i=1}^{N}$ follow the same
distribution and let the range of the surrogate statistic be $\mathbf{A}$,
while the support of the null distribution be $\mathbf{B}$, and $\mathbf{C}=%
\mathbf{B} \backslash \mathbf{A} $ be the complement set of $\mathbf{A}$
given $\mathbf{B}$. Then according to the rule, one rejects the null
hypothesis whenever the original statistic $s_{0}\in \mathbf{C}$, and the
actual false rejection rate is the probability $\Pr (s_{0}\in \mathbf{C}
|\{s_{i}\}_{i=1}^{N})$, which usually will not simply depend on the number
of surrogate realizations.

In this communication we will propose a new statistic, namely the Wilcoxon
signed rank statistic, for detection of the potential nonlinearity of a
scalar time series in the framework of Theiler \textit{et al. }\cite{theiler
testing}. This statistic could be derived from the linear prediction error
of the time series and proves to be a Wilcoxon variate under weak
conditions. As an advantage, it has a known null distribution, thus
inference with exact confidence level becomes possible based on the
knowledge of the statistic distribution. Furthermore, this statistic could
be applied to a wider range of linear stochastic process, including the
stationary Gaussian colored noise discussed in \cite{theiler testing}. We
will introduce the detail of the derivation of the Wilcoxon signed rank
statistic in Section 2. For demonstration, we will apply this statistic to
several examples in Section 3. Finally we will summarize the whole work in
Section 4.

\section{Methodology}

Now let us begin the introduction of the new statistic. As the first step,
we would like to specify the null hypothesis $H_{0}$ to be tested. Given a
stationary time series $\{x_{i}\}$ with finite coherence time \footnote{%
That is, its linear correlation will eventually tend to zero.} \cite{theiler
detecting}, our objective is to detect if there exists nonlinearity of the
underlying system. To perform a hypothesis test, we instead assume that the
time series $\{x_{i}\}$ is from a linear stochastic process with independent
jointly symmetric innovations, which is the null hypothesis $H_{0}$ to be
tested in our work. For the processes that are not consistent with $H_{0}$,
they will be attributed to the alternative null hypothesis $H_{1}$. For
clarity, let us explain with more detail (readers are also referred to the
original works \cite{theiler testing}). In general, the data generation
processes can be classified into linear or nonlinear stochastic processes
and linear or nonlinear deterministic systems. However, for linear
deterministic systems such as a sine wave, since they usually appear very
regular (their portraits in phase spaces are fixed points or limit cycles),
it is not difficult to detect the underlying mechanism. Thus in this work we
will exclude them from our discussions \footnote{%
A fixed point with disturbance can be considered as a random walk process,
which is within the scope of our consideration with the constraints imposed
in this work. For the cases of limit cycles contaminated with noise, their
coherence time will usually not be finite. For detection of such
trajectories, readers are referred to, e.g., \cite{small surrogate,luo
surrogate}}.

Also note that, theoretically it is possible that a linear stochastic
process has asymmetric innovation terms. However, linear stochastic
processes with symmetric innovations are often good approximations in many
practical situations, e.g., the Gaussian distributions of the driving force
of a Browning motion, the thermal noise in an electronic circuit, the
channel noise in telecommunication etc. Thus within the scope of our
discussion, when we reject the null hypothesis $H_{0}$, we will say that
there is a high possibility that the time series is generated from a
nonlinear system (either deterministic or stochastic) \footnote{%
Readers are referred to, e.g. \cite[p.~244-245]{galka topics}, for the
discussion on the formal interpretation of the test result.}. And this
result can be taken as the primitive step for further investigation of the
underlying system, for example, model building.

For any stationary linear stochastic process, it can be presented by an
autoregressive moving average (ARMA) process thanks to the Wold's
decomposition theorem. For convenience in the later discussion, whenever
feasible, we will use a $p$-th order autoregressive ($AR(p)$) processes
instead to describe a linear stochastic process with the concrete forms of 
\begin{equation}
x_{i}=a_{0}+\sum\nolimits_{j=1}^{p}a_{j}x_{i-j}+\epsilon _{i}\text{,}
\label{AR_P}
\end{equation}%
where $\epsilon _{i}$ denotes the innovation terms, which are assumed in our
null hypothesis to be independent of $x_{i}$, mutually independent of each
other and have a distribution with joint symmetry. By \textquotedblleft
joint symmetry\textquotedblright\ of a stochastic process $\{\epsilon _{i}\}$
we mean that there exists some constant $\mu $ so that $\{\epsilon _{i}-\mu
\}$ and $\{\mu -\epsilon _{i}\}$ have the same joint distributions, i.e. the
probability density function (PDF) $f(\epsilon _{1}-\mu ,\epsilon _{2}-\mu
,...)=f(\mu -\epsilon _{1},\mu -\epsilon _{2},...)$ \cite{dufour}. Clearly,
linear autocorrelated Gaussian processes examined in \cite{theiler testing}
are consistent with our null hypothesis. Yet the coverage could be extended
to a wider range, e.g. the linear stationary processes with independent (not
necessarily identical) and jointly symmetric innovations.

With the above null hypothesis, we then need to choose a discriminating
statistic to determine whether to reject the null or not. To derive our
statistic, let us first consider the problem of predicting $k$-step ahead
value of a linear stochastic process $\{x_{i}\}$ with jointly symmetric
innovations $\{\epsilon _{i}\}$. Let $\hat{x}_{i}^{k}$ be the prediction at
time $i$ ($\hat{x}_{i}^{k}=x_{i+k}$ if $k\leqslant 0$), and $%
e_{i}^{k}=x_{i+k}-\hat{x}_{i}^{k}$ denote the corresponding prediction
error. For the purpose of prediction, we choose the ordinary least square
(OLS) linear predictor $\hat{x}_{i}^{k}=a_{i,0}+\sum\nolimits_{j=1}^{p^{%
\prime }}a_{i,j}\hat{x}_{i}^{k-j}$ \footnote{%
Here $p^{\prime }$ is the specified fitting order, and $%
\{a_{i,j}:j=0,1,...,p^{\prime }\}$ are the estimated parameters at time $i$
via the criterion of ordinary least squares, which aims to minimize the
forward squared error of the prediction.} with $a_{i,j}$ being the the $%
(j+1) $th coefficient of the OLS predictor, estimated based on the history $%
\{x_{i},x_{i-1},x_{i-2},...\}$. In general situations, the OLS predictor may
not perform as well as other procedures like the forward/backward
least-square algorithm, however, it does possess an interesting property
that might not be shared by other algorithms. The fact is that, as proved in 
\cite{dufour}, when an OLS linear predictor is used to predict the $k$-step
ahead value of a linear stochastic processes $\{x_{i}\}$ with jointly
symmetric innovations, even if the fitting order of the predictor is
misspecified (either lower or higher), and thus inaccurate estimated
parameters are adopted for prediction, the distributions of the prediction
error $e_{i}^{k}$ will still be symmetric about zero, i.e. $e_{i}^{k}$ and $%
-e_{i}^{k}$ share the same distribution. This fact immediately implies that
the probability $\Pr (e_{i}^{k}>0)=\Pr (-e_{i}^{k}>0)=\Pr (e_{i}^{k}<0)=1/2$
when the distribution of $e_{i}^{k}$ is continuous so that the probability $%
\Pr (e_{i}^{k}=0)=0$ in the sense of Lebesgue measure (see \cite{luger exact}
and 
\cite[Lemma 10.1.24]{randles introduction}
for more details).

Let $\{I_{i}\}_{i=1}^{m}$ be an indicator series with $m$ data points so
that $I_{i}(e_{i}^{k})=1$ if $e_{i}^{k}>0$ and $I_{i}(e_{i}^{k})=-1$ if $%
e_{i}^{k}<0$. Clearly $I_{i}$ is a Bernoulli variate uniformly distributed
on $\{-1,1\}$, i.e. $\Pr (I_{i}=-1)=\Pr (I_{i}=1)=1/2$. With this knowledge,
we can derive a class of linear signed rank statistics 
\begin{equation}
SR_{m}=\sum\nolimits_{i=1}^{m}I_{i}(e_{i}^{k})\times S_{i}(rank(|e_{i}^{k}|))
\label{signed rank}
\end{equation}%
to test our null hypothesis, where $\{S_{i}(\cdot )\}$ is the set of scores
of the series $\{|e_{i}^{k}|\}_{i=1}^{m}$ with $rank(|e_{i}^{k}|)$ denoting
the rank (in the ascending order) of the absolute value $|e_{i}^{k}|$ among $%
\{|e_{i}^{k}|\}_{i=1}^{m}$ 
\cite[p. 252]{randles introduction}%
\footnote{%
Since $e_{i}^{k}$ are symmetric about zero, the absolute value of $e_{i}^{k}$
is adopted to remove the possible dependence between $S_{i}$ and $I_{i}$.}.
Here we choose $S_{i}(rank(|e_{i}^{k}|))=rank(|e_{i}^{k}|)$ so that 
\begin{equation}
SR_{m}=\sum\nolimits_{i=1}^{m}i\times I_{i}(e_{i}^{k})  \label{wilcoxon}
\end{equation}%
is the widely used Wilcoxon signed rank statistic, which is discretely
distributed. One could obtain the full knowledge of the distribution by
enumerating all of its possible values, however, this is an ineffective way
especially when $m$ is large. A remedy to this problem is to use the
Gaussian distribution $N(0,m(m+1)(2m\,+1)/6)$ for approximation based on the
limit central theorem. In \cite[chaper 2]{maritz distribution} it is shown
that the approximation works well even for small numbers, say, $m=6$.
Therefore in this work we will adopt the strategy of distribution
approximation.

Since we know the distribution of the test statistic, based on the
realization value of $SR_{m}$, we can determine whether to reject the null
hypothesis with an exact confidence level. Take two-sided test as an
example, if we want the type-I error (the false rejection rate of a correct
null) to be less than $\alpha $, then we first find two critical values $%
n_{u}$ and $n_{l}$ such that $n_{u}$ is the largest integer satisfying $\Pr
(SR_{m}>n_{u})<\alpha /2$ and $n_{l}$ is the smallest integer satisfying $%
\Pr (SR_{m}<n_{l})<\alpha /2$. If for a time series in test, its statistic $%
SR_{m}>n_{u}$ or $SR_{m}<n_{l}$, then we reject the null hypothesis. The
false rejection rate is $\alpha $, or in other words, the confidence level
to reject the null hypothesis is $1-\alpha $. The procedures to perform
one-sided tests are similar, except that we need to locate only one critical
value, $n_{u}$ or $n_{l}$, which instead satisfies $\Pr
(SR_{m}>n_{u})<\alpha $ or $\Pr (SR_{m}<n_{l})<\alpha $ separately for right
or left side test.

For a linear stochastic process consistent with our null hypothesis, in
principle one would expect that the actual rejection rate will be the same
as the nominal (pre-specified) one. However, for a nonlinear system
violating the hypothesis, it would generally engender higher rejection rates
because of the asymmetry of the prediction errors (especially for the
surrogate data, see the discussion in the next section), which is the
essential idea inside our method.

\section{Numerical results}

We will apply the above idea to test our null hypothesis for nonlinearity
detection. The whole procedures go as follows: For each time series $%
\{x_{i}\}$ in test, we first predict its one-step ahead values $\{\hat{x}%
_{i}^{1}\}$ via the OLS linear predictor and calculate the prediction error 
\footnote{%
Theoretically any $k$ shall be okay since one always has that $\Pr
(e_{i}^{k}>0)=\Pr (-e_{i}^{k}>0)=\Pr (e_{i}^{k}<0)=1/2$. However, we do
recommend the choice of $k=1$ because in practice there might be additional
noise from computer (e.g round-off effect) based on the OLS predictor if $%
k\geq 2$, which may affect the symmetry of the prediction error.}. Suppose
the error series $\{e_{i}^{1}\}_{i=1}^{m}$ has $m$ data points, then we use
the normal distribution $N(0,m(m+1)(2m\,+1)/6)$ to find the critical values (%
$n_{u}$ and $n_{l})$ for two-sided test at the nominal confidence level of $%
95\%$. If the calculated Wilcoxon statistic $SR_{m}\notin $ $[n_{l}$, $%
n_{u}] $, then we can reject the null hypothesis with the false rejection
rate of $5\%$. In the following we will demonstrate through several examples
the power of our scheme for the null hypothesis test.

\begin{figure}[tbp]
\centering
\par
\includegraphics[width=5in]{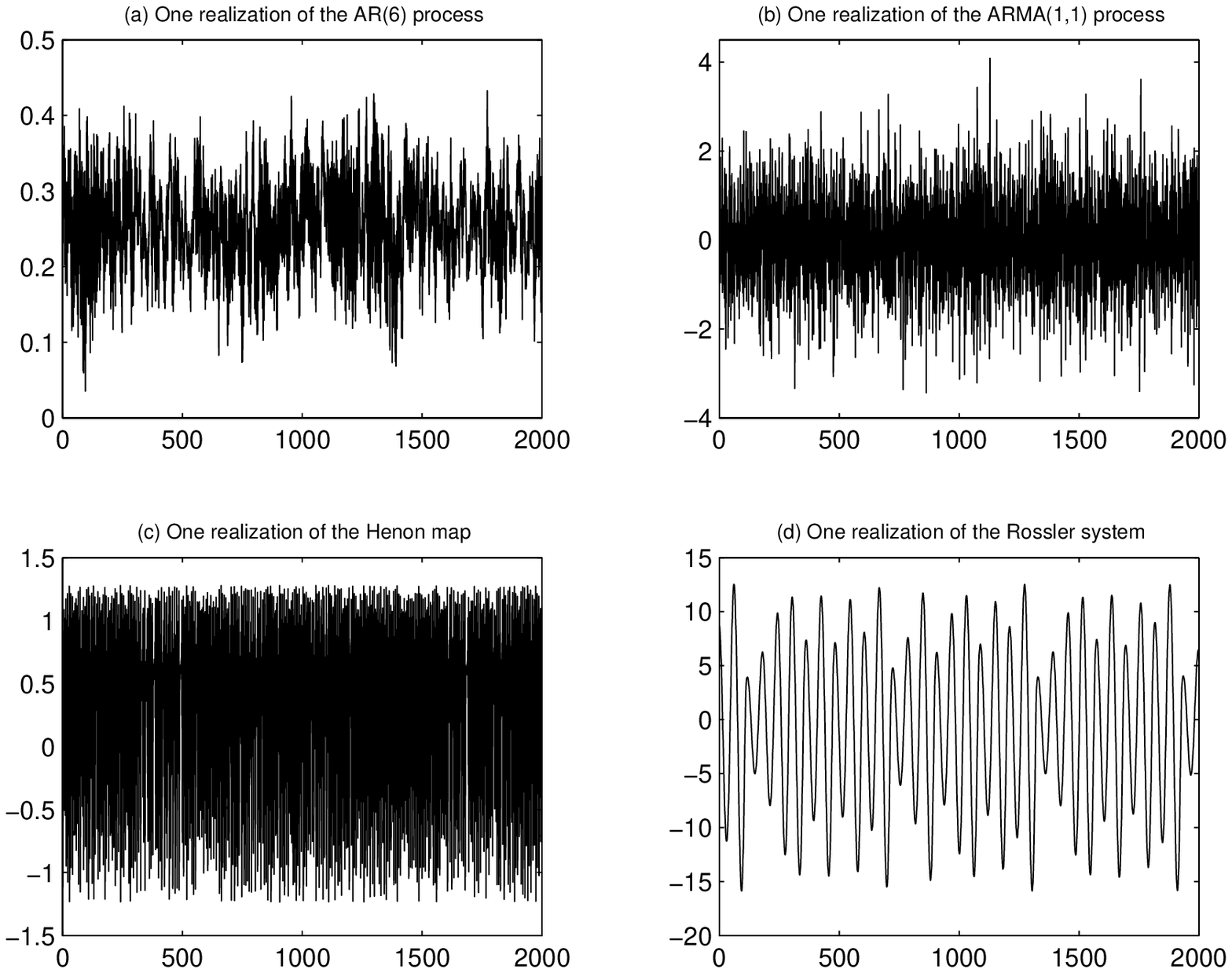}
\caption{Waveform of a realization of: (a) the $AR(6)$ process, (b) the $%
ARMA(1,1)$ process, (c) the Henon map, and (d) the R\"{o}ssler system.}
\label{DGPplots}
\end{figure}

The first example is an $AR(6)$ process $x_{i}=\sum%
\nolimits_{j=1}^{6}a_{j}x_{i-j}+\epsilon _{i}$ with coefficients $%
(a_{1},...,a_{6})=$ $(0.6$, $0$, $0.5$, $0$, $-0.6$, $0.3)$, where
innovations $\{\epsilon _{i}\}$ are uniformly distributed on interval $[0$, $%
0.1]$ (symmetric about $0.05$). The second is an $ARMA(1,1)$ process, i.e. $%
x_{i}=a_{1}x_{i-1}+\epsilon _{i}-b_{1}\epsilon _{i-1}$ with parameters $%
a_{1}=b_{1}=0.5$, where innovation terms $\{\epsilon _{i}\}$ follow the
normal distribution $N(0,1)$. The third data generation process (DGP) is the
H\'{e}non map \cite{henon} $H(x,y)=(y+1-\alpha x^{2},0.3x)$, where parameter 
$\alpha $ is uniformly drawn from the interval $[1.35$, $1.4]$. We will take
out the first coordinate $x$ for test. The final case is the R\"{o}ssler
system \cite{rossler} with continuous description equations of $(\dot{x},%
\dot{y},\dot{z})=(-y-z,x+0.15y,0.2+xz-cz)$, where parameter $c$ is uniformly
drawn from the interval $[9.5$, $10]$. The sampling time is $0.1$ time
units, and the observations for calculation are taken from the second
coordinate $y$. The waveforms of the realizations of each DGP are plotted in
Fig. \ref{DGPplots}.

Since usually one does not know the true order of an underlying process, a
fitting order has to be specified for prediction. For this purpose, one may
adopt the Akaike or Schwarz information criterion \cite{Burnham model}.
However, as aforementioned, if the underlying process of the test time
series is linear stochastic with jointly symmetric innovations,
misidentification of the fitting order would also lead to the symmetric
prediction error. Thus we need not seek the optimal fitting order for our
prediction. Instead, we could simply choose several fitting orders for all
of the DGPs, say, those starting from $6$ to $10$. To indicate the power of
our test scheme, for all of the DGPs in examination, we produce $1000$
realizations with $2000$ data points for each, and predict the one-step
ahead values for the last $500$ data points. For demonstration, the
prediction errors with $p^\prime=6$ are illustrated in Fig. \ref%
{DGPErrorplots} for each DGP (The results with other fitting orders are
similar and thus not reported here). With the prediction errors, we
calculate the Wilcoxon signed rank statistic to determine whether to reject
the null or not. We record the rejection numbers of our null hypothesis and
indicate them in Table \ref{table1}. For the $AR(6)$ and $ARMA(1,1)$
processes, the rejection rates are nearly $5\%$, as we expect. For the R\"{o}%
ssler system, although the rejection rates are various for different fitting
orders, in general they are much higher than $5\%$, thus we can reject the
null hypothesis. While for the Henon map, we see that the rejection rates
are only slightly higher than the nominal one. The explanation may be that,
in principle there is no universal scores in Eq. (\ref{signed rank}) that
bring the most powerful signed rank statistic for all systems (\cite[Theorem
10.1.19.]{randles introduction}), and the sensitivity of this class of
statistics might dramatically decrease for certain systems. The problem,
however, could be relieved to some extent when we apply our framework to
surrogate tests, as to be shown below.

In practical situations, one often has only a scalar time series on hand.
Therefore, for the reliability of the test, we suggest that one first uses
the bootstrap method, such as \cite{theiler testing} and \cite{nakamura}, to
generate a number of surrogates, and then calculates the test statistic of
the surrogates and determines whether to reject the null hypothesis or not.
If the actual rejection rate is higher than the nominal one (i.e. the rate
when the original time series is consistent with our null hypothesis), then
we can safely reject the null hypothesis. For illustration, let us examine
the previous examples again. For each example, we generate only one sample,
and use the bootstrap method to generate $1000$ of its surrogates. In our
test scheme, we adopt the temporal-shift algorithm in \cite{nakamura,luo
surrogate} to generate surrogates since this algorithm does not require the
Fourier transform and thus avoids some of its shortcomings \footnote{%
A good review of the conventional surrogate algorithms can be found in, for
example, \cite[chapter 11]{galka topics}}. The main idea of the algorithm is
that, if a time series $\{x_{i}\}$ is linear stochastic with the form of $%
x_{i}=a_{0}+\sum\nolimits_{j=1}^{p}a_{j}x_{i-j}+\epsilon _{i}$, then for any
coefficients $\beta $ and $\gamma $, the surrogates $\{y_{i}^{\tau }=\beta
x_{i}+\gamma x_{i+\tau }\}$ also follow linear stochastic forms with the
constants of $(\beta +\gamma )a_{0}$ and the innovations of $\{\beta
\epsilon _{i}+\gamma \epsilon _{i+\tau }\}$ respectively. A linear
stochastic time series $\{x_{i}\}$ will always produce linear stochastic
surrogates $\{y_{i}^{\tau }=\beta x_{i}+\gamma x_{i+\tau }\}$, therefore in
principle we cannot reject the null hypothesis test via the Wilcoxon
statistic. Following the suggestion in \cite{luo surrogate}, in our
calculations we let parameter $\beta $ be uniformly drawn from the interval $%
[0.6$, $0.8]$ and parameter $\gamma =(1-\beta ^{2})^{1/2}$ to produce
surrogates. The results of null hypothesis test are presented in Table \ref%
{table2}, which appear different from those in Table \ref{table1}. For the
two linear stochastic processes, the rejection numbers decrease to zero.
This is because that the original samples of the linear stochastic processes
are not rejected in the tests (as $95\%$ of the samples are), therefore all
its surrogates, which are essentially the addition of two segments of the
original samples, will also not be rejected in the tests. However, the
situation for nonlinear systems is different. Adding together two segments
of the original samples from nonlinear systems increases the complexity,
thus the rejection rates of the surrogates increase correspondingly compared
to those in Table \ref{table1}.

For comparison, we also generate $1000$ surrogates through the
constrained-realization method \cite{theiler testing}, which preserves the
linear correlation of the original time series but introduce some
randomization to the phase of the Fourier transform. The computation results
are indicated in Table \ref{table3}, from which we could see that, the
rejection rates of linear cases drop as well in contrast to those in Table %
\ref{table1}. However, the rates of nonlinear cases remain close to those in
Table \ref{table1}, this is because, with the preservation of the linear
correlation, the phase perturbation of the surrogates does not significantly
affect the data structure, thus the rejection rates of the nonlinear systems
are close to those in Table \ref{table1}.

An further question is that: what will happen if the data in test is linear
but with asymmetric innovation terms? To answer this question, let us see an
example. We construct the following process 
\begin{equation}
x_{i}=0.1+0.8x_{i-1}-0.5x_{i-2}+0.2x_{i-3}+\epsilon _{i}  \label{betanoise}
\end{equation}%
with the innovation term $\epsilon _{i}$ following the $beta(2,5)$
distribution. Here 
\begin{equation}
beta(\alpha ,\beta )\equiv \frac{1}{B(\alpha ,\beta )}s^{\alpha
-1}(1-s)^{\beta -1}
\end{equation}%
with $B(\alpha ,\beta )$ denoting the beta function. We choose the same
computation settings as previously adopted. The results are summarized as
follows,

\begin{enumerate}
\item For 1000 realizations generated by Eq. (\ref{betanoise}), the
rejection numbers corresponding to the fitting orders ($6$ to $10$) are $98
, 99, 103,105,109$ respectively.

\item For 1000 surrogates produced by the temporal shift algorithm based on
one realization of Eq. (\ref{betanoise}), the rejection numbers
corresponding to fitting orders are $0, 0, 0, 0, 0$ respectively.

\item For 1000 surrogates produced by the conventional constrained
realization algorithm, the rejection numbers are $109, 97, 100, 117, 129$
respectively.
\end{enumerate}

Based on the results, one can see that in $(3)$ one can correctly reject the
null hypothesis but in $(2)$ one fails. Because the temporal shift algorithm
is simply based on the linear superposition principle, although failing to
reject the null hypothesis in $(2)$, it does tell us some useful
information, i.e., the data in test is unlikely to be nonlinear (otherwise
the rejection rate might obviously exceed the nominal, 5\%, because of the
superposition of the surrogates). In contrast, rejecting the null hypothesis
in $(3)$ says that the data in test cannot be generated by a linear process
with jointly symmetric innovation terms. Thus conbining the results in $(2)$
and $(3)$ might shed light on the underlying process, which is to be
investigated in a future work.

\begin{figure}[tbp]
\centering
\par
\includegraphics[width=5in]{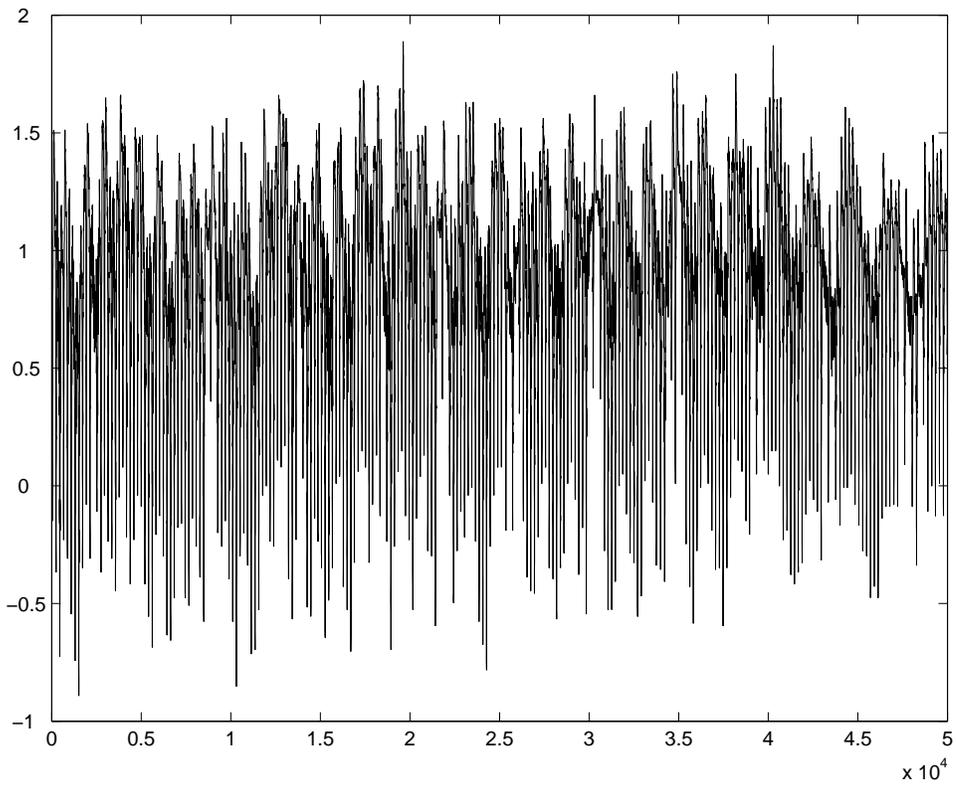}
\caption{Waveform of a human electrocardiogram (ECG) record during
ventricular fibrillation (VF)}
\label{vf}
\end{figure}

As an application, we apply our scheme to examine an electrocardiogram (ECG)
record during ventricular fibrillation (VF), a segment of which is plotted
in Fig. \ref{vf} (for the details of data acquisition, see \cite{small
automatic}). The segment in examination has $50,000$ data points, thus we
use the temporal-shift method to generate $1000$ surrogates since it needs
less computations than the constrained-realization algorithm. We also adopt
the same fitting orders, from $6$ to $10$, to predict $100$ one-step ahead
values. As the results, the correspondingly rejections of our null
hypothesis are $566,$ $541,$ $530,$ $550,$ and $521$ (all out of $1000$
tests), a strong hint of nonlinearity.

\section{Conclusion}

To summarize, we have proposed an exact nonparametric inference scheme to
detect the potential nonlinearity in a scalar time series. The exactness of
our inference comes from the knowledge of the exact distribution of the
adopted discriminating statistic, i.e., the Wilcoxon signed rank statistic,
which indicates remarkable test power through the several examples examined
in this communication. The advantages of this statistic include: it
possesses the known null distribution. Thus it is easy for us to find the
exact confidence interval for the inference of the underlying process.
Furthermore, the exactness of the statistic distribution does not rely on
the size of the sample time series in test. Comparatively, for many
nonlinear discriminating statistics adopted in the literature, e.g. the
correlation dimension, computations with too short samples may cause serious
distortions.

XL was supported by a Hong Kong University Grants Council Competitive
Earmarked Research Grant (CERG) number PolyU 5216/04E.

\clearpage

\clearpage

\begin{figure}[tbp]
\centering
\par
\includegraphics[width=5in]{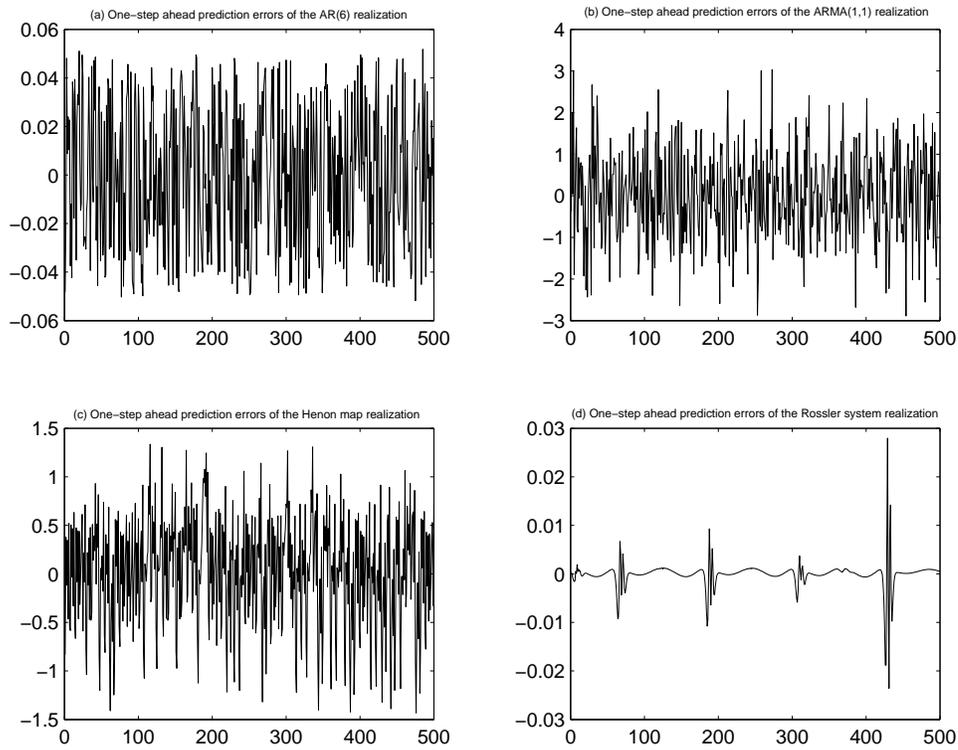}
\caption{One-step ahead prediction errors ($p^\prime=6$) of a realization
of: (a) the $AR(6)$ process, (b) the $ARMA(1,1)$ process, (c) the Henon map,
and (d) the R\"{o}ssler system.}
\label{DGPErrorplots}
\end{figure}

\clearpage

\clearpage

\begin{table}[!h]
\centering 
\parbox{4in}{\caption{\label{table1} Numbers of rejections of the null hypothesis
(out of 1000 replica) for data generation processes (DGPs) with different fitting orders.
}} 
\begin{tabular*}{4in}{cccccc}
\hline\hline
DGP & \multicolumn{5}{c}{Rejections for the fitting order of} \\ \cline{2-6}
& 6 & 7 & 8 & 9 & 10 \\ \hline
AR(6) & 55 & 54 & 54 & 53 & 50 \\ 
ARMA(1,1) & 44 & 41 & 43 & 44 & 48 \\ 
Henon & 76 & 55 & 60 & 62 & 67 \\ 
R\"{o}ssler & 990 & 442 & 752 & 298 & 458 \\ \hline\hline
\end{tabular*}%
\end{table}

\clearpage

\begin{table}[!h]
\centering 
\parbox{4in}{\caption{\label{table2} Rejections of the null hypothesis
for $1000$ temporal-shift surrogates of DGPs with different fitting orders.
}} 
\begin{tabular*}{4in}{cccccc}
\hline\hline
DGP & \multicolumn{5}{c}{Rejections for the fitting order of} \\ \cline{2-6}
& 6 & 7 & 8 & 9 & 10 \\ \hline
AR(6) & 0 & 0 & 0 & 0 & 0 \\ 
ARMA(1,1) & 0 & 0 & 0 & 0 & 0 \\ 
Henon & 432 & 307 & 308 & 203 & 207 \\ 
R\"{o}ssler & 1000 & 553 & 1000 & 593 & 626 \\ \hline\hline
\end{tabular*}%
\end{table}

\clearpage

\begin{table}[!h]
\centering 
\parbox{4in}{\caption{\label{table3} Rejections of the null hypothesis
for $1000$ constrained-realization surrogates of DGPs with different fitting orders.
}} 
\begin{tabular*}{4in}{cccccc}
\hline\hline
DGP & \multicolumn{5}{c}{Rejections for the fitting order of} \\ \cline{2-6}
& 6 & 7 & 8 & 9 & 10 \\ \hline
AR(6) & 14 & 14 & 10 & 11 & 9 \\ 
ARMA(1,1) & 1 & 1 & 2 & 1 & 2 \\ 
Henon & 83 & 56 & 63 & 61 & 83 \\ 
R\"{o}ssler & 992 & 449 & 753 & 301 & 460 \\ \hline\hline
\end{tabular*}%
\end{table}

\end{document}